\documentclass[aps,prl,twocolumn,groupedaddress,showpacs]{revtex4}

\usepackage{amsmath}
\usepackage{amssymb} 
\usepackage{graphics} 
\usepackage{epsfig}
\usepackage{color}

\allowdisplaybreaks[2]

\newcommand{\beq}{\begin{equation}}
\newcommand{\eeq}{\end{equation}}
\newcommand{\beqa}{\begin{eqnarray}}
\newcommand{\eeqa}{\end{eqnarray}}
\newcommand{\bea}{\begin{eqnarray}}
\newcommand{\eea}{\end{eqnarray}}
\newcommand{\ba}{\begin{align}}
\newcommand{\ea}{\end{align}}
\newcommand{\dst}{\displaystyle}
\newcommand{\non}{\nonumber}

\newcommand{\TRUE} {\textrm{\sc true}}
\newcommand{\FALSE}{\textrm{\sc false}}
\newcommand{\T}    {\textrm{\sc t}}
\newcommand{\F}    {\textrm{\sc f}}
\newcommand{\Q}    {\mathbf{Q}}

\newcommand{\Qc}   {{\cal Q}}

\newcommand{\nb}   {\mathbf{n}}

\newcommand{\n}    {n}

\newcommand{\ku}   {{\underline{k}}}

\newcommand{\x}    {{\mathbf{x}}}

\newcommand{\xb}   {{\mathbf{x}}}
\newcommand{\Xb}   {{\mathbf{X}}}

\newcommand{\zb}   {{\mathbf{z}}}
\newcommand{\Zb}   {{\mathbf{Z}}}

\newcommand{\AbL}  {{\mathbf{A}_L}}
\newcommand{\Bb}   {{\mathbf{B}}}
\newcommand{\BbL}  {{\mathbf{B}_L}}

\newcommand{\Rb}   {{\mathbf{R}}}
\newcommand{\Rbt}  {\widetilde{\mathbf{R}}}
\newcommand{\eb}   {{\mathbf{e}}}
\newcommand{\Eb}   {{\mathbf{E}}}
\newcommand{\EbhL} {{\hat{\mathbf{E}}_L}}
\newcommand{\Eh}   {{\hat{E}}}
\newcommand{\0}    {\mathbf{0}}
\newcommand{\1}    {\mathbf{1}}
\newcommand{\Nb}   {{\mathbf{N}}}

\newcommand{\CbL}  {\mathbf{C}_L}
\newcommand{\DbL}  {\mathbf{D}_L}
\newcommand{\GbL}  {\mathbf{G}_L}
\newcommand{\Sb}   {\mathbf{S}}

\newcommand{\Cc}   {{\cal C}}

\newcommand{\TbL}  {\mathbf{T}_L}
\newcommand{\Yb}   {\mathbf{Y}}
\newcommand{\Mb}   {\mathbf{M}}
\newcommand{\Mt}   {\widetilde{M}}
\newcommand{\Mbt}  {\widetilde{\mathbf{M}}}
\DeclareMathOperator{\Tr}{Tr}

\newcommand{\phiC} {\phi_L^\textrm{\scriptsize C}}
\newcommand{\phiI} {\phi_L^\textrm{\scriptsize I}}

\newcommand{\vmu}  {{\vec{\mu}}}
\newcommand{\vnu}  {{\vec{\nu}}}
\newcommand{\vkappa}{{\vec{\kappa}}}
\newcommand{\vlambda}{{\vec{\lambda}}}

\newcommand{\CL}[1][L]{\langle C_{#1}\rangle_N}
\newcommand{\OL}[1][L]{\langle\Omega_{#1}\rangle_N}

\newcommand{\OLinf}[1][L]{\langle\Omega_{#1}\rangle_\infty}
\newcommand{\rC} {r^\textrm{\scriptsize C}}
\newcommand{\rI} {r^\textrm{\scriptsize I}}

\newcommand{\dr} {\Delta r}

\newcommand{\Nat}{\mathbb{N}}

\newcommand{\trm}{\textrm}

\begin{document}

\preprint{LU TP 02-43}

\title{Counting attractors in synchronously updated random Boolean networks}


\author{Bj\"orn Samuelsson}
\email[]{bjorn@thep.lu.se}
\author{Carl Troein}
\email[]{carl@thep.lu.se}

\affiliation{
Complex Systems Division, Department of Theoretical Physics\\
Lund University,  S\"olvegatan 14A,  S-223 62 Lund, Sweden}


\date{2004-12-16}

\begin{abstract}

Despite their apparent simplicity, random Boolean networks
display a rich variety of dynamical behaviors. Much work has been
focused on the properties and abundance of attractors. We here
derive an expression for the number of attractors in the
special case of one input per node. Approximating
some other non-chaotic networks to be of this class, we apply
the analytic results to them. For this approximation, we observe a
strikingly good agreement on the numbers of attractors of various lengths.
Furthermore, we find that for long cycle lengths, there
are some properties that seem strange in comparison to real
dynamical systems. However, those properties can be interesting
from the viewpoint of constraint satisfaction problems.

\end{abstract}

\pacs{89.75.Hc, 02.10.Ox}

\maketitle


\section{Introduction}

Random Boolean networks have long enjoyed the attention of
researchers, both in their own right and as simplistic models,
in particular for gene regulatory networks. The
properties of these networks have been studied for a variety
of network architectures, distributions of Boolean rules,
and even for different updating strategies. The simplest and most
commonly used strategy is to synchronously update all nodes.
Networks of this kind have been investigated
extensively, see e.g.\ \cite{Kauffman:69, Derrida:86a,
Bilke:01, Aldana:03, Socolar:03}.

The networks we consider are, generally speaking, such where the
inputs to each node are chosen randomly with equal probability among
all nodes, and where the Boolean rules of the nodes are picked
randomly and independently from some distribution. The network
state is updated synchronously, with the state of a node at any
time being a function of the state of its inputs at the previous
time step.

In this work we determine analytically the numbers of attractors of
different lengths in networks with connectivity (in-degree) one. We
compare these results to networks of higher connectivity and find a
remarkable degree of agreement, meaning that networks of single-input nodes
can be employed to approximate more complicated networks, even for
small systems. For large networks, a reasonable level of correspondence
is expected. See \cite{Bastolla:98} on effective connectivity for
critical networks, and \cite{Kauffman:04} on the limiting numbers of
cycles in subcritical networks.

Random Boolean networks with connectivity one have been investigated
analytically in earlier work \cite{Flyvbjerg:88, Drossel:04}. In those
papers, a graph-theoretical approach was employed. In this work,
we use another approach \cite{Samuelsson:03} which is based on the
ideas of fixed point analysis. Our approach is a powerful tool for counting
the average numbers of attractors, but it does not give any direct
information on the properties of the attractor basins.

For long cycles, especially in large networks, there are some
artefacts that make comparisons to real networks difficult. For
example, the integer divisibility of the cycle length is
important, see e.g.\ \cite{Flyvbjerg:88, Bastolla:98, Samuelsson:03,
Kauffman:04, VKaufman:04}. Also, the total number of attractors grows
superpolynomially with system size in critical networks
\cite{Samuelsson:03, Drossel:04}, and most of the attractors have tiny
attractor basins as compared to the full state space. In this work,
such artefacts become particularly apparent, and we think that long
cycles are hard to connect to real dynamical systems.

Comparisons to real dynamical systems, on the other hand, still seem to
be relevant with regard to fixed points and some stability properties
\cite{Kauffman:03, Kauffman:04}. An interesting way to
make more realistic comparisons regarding cycles is to consider those
attractors that are stable with respect to repeated infinitesimal
changes in the timing of updating events \cite{Klemm:04}.

Furthermore, the problem of finding attractors with small attractor
basins, for a given network, can be seen as a hard constraint
satisfaction problem. Those properties that are artefacts when
comparing to dynamical systems, could be keys to the understanding of
real life combinatorial optimization problems.


Throughout this paper, $N$ denotes the number of nodes in the network,
and $L$ the length of an attractor, be it a cycle ($L > 1$)
or a fixed point ($L=1$). For brevity we use the term
{\it $L$-cycle}, and understand this to mean an attractor
such that taking $L$ time steps forward produces the initial
state. When $L$ is the smallest positive integer fulfilling this,
we speak of a {\it proper $L$-cycle}.
We denote the number of proper $L$-cycles with $C_L$. The
average over networks of a certain size is $\CL$, so the
average number of network states that are part of a proper
$L$-cycle is $L\CL$. Related to this is $\OL$, the number
of states that reappear after $L$ time steps and hence
are part of any $L$-cycle, proper or not.


To calculate the average number of attractors, we use the same
approach as in \cite{Samuelsson:03}, whereby we first transform
the problem of finding an $L$-cycle into a fixed point problem, and
then find a mathematical expression for the average number of solutions
to that problem. 

In the case that every node has one input, $\OL$ can be calculated
analytically for any $N$. The limit $\OLinf$ has a relatively simple
form, and this limit can be
mapped to the $1$-input case for any subcritical network of multi-input nodes,
as shown in \cite{Kauffman:04}. Furthermore,
critical networks of multi-input nodes seem to show strong similarities to
the corresponding networks of $1$-input nodes.

\section{Theory}

Assume that a Boolean network performs a proper $L$-cycle. Then, each node
performs one of $m\equiv2^L$ series of output values. We call these
{\it $L$-cycle series}. We associate each $L$-cycle series with an index
$i$, such that $i\in\{0,1,\ldots,m-1\}$. For convenience, we let the
constant $L$-cycle series have the indicies $0$ and $1$, in such a way that
a constant $\FALSE$ output has index $0$ whereas a constant $\TRUE$ output
has index $1$.

If we view each $L$-cycle
series as a state, an $L$-cycle of the entire network turns into a fixed
point (in this enlarged state space).
$L\CL$ is then the average number of input
states (choices of $L$-cycle series), for the whole network, such that
the output is the same as the input. 

It is inconvenient to work directly with $\CL$. Let $\OL$ denote the
average number of states that reappear after $L$ timesteps.
Such a state is part of a proper $\ell$-cycle where $\ell$ is a
divisor to $L$. $\CL$ can be calculated from $\OL$, using the set
theoretic inclusion--exclusion principle.

\subsection{Expressing $\CL$ in terms of $\OL$}

Let ${\cal C}_L$ denote the set of network state sequences that
represent proper $L$-cycles. Similarly, let $\omega_L$ denote the
non-proper counterpart to $\Cc_L$, meaning that $\omega_L =
\bigcup_{\ell|L} {\cal C}_\ell$ where $\ell|L$ means that $\ell$ divides
$L$.

Consider a given network, and let $\Qc$ denote the set of sequences of
states, $\Q$, that are consistent with the network dynamics. Let
$\hat{\omega}_L = \omega_L \cap \Qc$ and $\hat{\Cc}_L = \Cc_L \cap
\Qc$. Then, $\Omega_L=\lvert\hat{\omega}_L\rvert$, and the number of
proper $L$-cycles of that network is given by
$C_L=\frac1L\lvert\hat{\Cc}_L\rvert$. To calculate $C_L$, we start by
expressing $\hat{\Cc}_L$ in terms of $\hat{\omega}_\ell$. We see that
\beq
  \hat{\Cc}_L = \hat{\omega}_L \,\big\backslash\!
        \bigcup_{\substack{1\leq \ell<L\\\ell|L}}\!\! \hat{\omega}_\ell
       = \hat{\omega}_L \,\big\backslash\!\! 
        \bigcup_{\substack{d\textrm{ prime}\\d|L}}\!\! \hat{\omega}_{L/d}
  \label{eq: proper states},
\eeq
because any positive $\ell$ dividing $L$ is also a divisor to a number of
the form $L/d$, where $d$ is a prime.

Then, the inclusion--exclusion principle, applied to
eq.~\eqref{eq: proper states}, yields
\beq
  C_L = \tfrac1L\hspace{-8pt}\sum_{\mathbf{s}\in\lbrace0,1\rbrace^{\eta_L}}\!\!
         (-1)^s\Omega_{L/d_L(\mathbf{s})},
\eeq
where $s = \sum_{i=1}^{\eta_{L}} s_i$, $d_{L}(\mathbf{s}) = \prod_{i=1}^{\eta_L}
(d_L^i)^{s_i}$ and $d_L^1,\ldots,d_L^{\eta_L}$ are the prime divisors to
$L$. For averages over randomly chosen $N$-node networks, we get
\beq
  \CL = \tfrac1L\hspace{-10pt}\sum_{\mathbf{s}\in\lbrace0,1\rbrace^{\eta_L}}\!\!
  (-1)^s\langle\Omega_{L/d_L(\mathbf{s})}\rangle_N~.
\label{eq: CL of OL}
\eeq

\subsection{Basic expressions for $\OL$}

Consider what a rule does when it is subjected
to $L$-cycle series on its inputs. It performs some Boolean
operation, but it also delays the output, giving a one step difference
in phase for the output $L$-cycle series. 
Let $A_L^i(\x)$ denote the
probability that a randomly selected rule will output $L$-cycle series
$i$, given that the input series are selected from the distribution
$\x = (x_0,\ldots,x_{m-1})$. Then, we get
\beq
  \OL = \!\sum_{\substack{\nb \in \Nat^m\\n=N}}\binom{N}{\nb}\prod_{i=0}^{m-1}
      [A_L^i(\nb/N)]^{\n_i}~,
\label{eq: OL}
\eeq
where $n=n_0+\cdots+n_{m-1}$. This is the same expression as presented in
\cite{Kauffman:04}.

Let $\eb_0$ be the vector with elements $e_0^i=\delta_{i0}$,
where $\delta$ is the Kronecker delta.
For networks of 1-input nodes, $A_L^i(\xb)$ is affine and can be
expressed as
\begin{align}
  A_L^i(\xb) &= (\x - \eb_0)\cdot\nabla A_L^i
               + A_L^i(\eb_0)\\
       &= \x\cdot\nabla A_L^i + \delta_{i0}c_0 + \delta_{i1}c_1~,
\end{align}
where $\rC$ and $\rI$ are the fractions of nodes that
copy and invert their input, respectively, while $c_0$ and $c_1$ are the
fractions of constant nodes that output $\FALSE$ and $\TRUE$, respectively. The
elements $\partial_jA_L^i$ of the gradient $\nabla A_L^i$ are nonzero
only if the $L$-cycle series $i$ can be the output of a node
receiving the $L$-cycle series $j$ on its input. This is true if the
series $j$ is the series $i$, or the inverse of $i$, rotated one step
backwards in time. Let $\phiC(i)$ and $\phiI(i)$, respectively, denote
those values of $j$. Then,
\beq
  \x\cdot\nabla A_L^i = \rC x_{\phiC(i)}+\rI x_{\phiI(i)}~.
  \label{eq: OLinf1g}
\eeq

Let $\BbL$ be an $m\times m$ matrix with elements
\beq
  B_L^{ij} = \partial_jA_L^i + \delta_{i0}(c_0 - 1) + c_1\delta_{i1}~.
\eeq
Then,
\beq
  \AbL(\xb) = \BbL\x + \eb_0
\label{eq: AbL}
\eeq
for all $\x$ such that $x_0 + \ldots + x_{m-1}=1$. Furthermore,
\beq
  \sum_{i=0}^{m-1}(\BbL\x + \eb_0)_i = 1
\eeq
for all $\x$. This property will be important later on.

From the theory of linear algebra, we know that any matrix can be
transformed to a triangular matrix. Thus, $\BbL$ can be rewritten as
\beq
  \BbL=\CbL^{-1}\DbL\CbL~,
\label{eq: BbL triangle}
\eeq
where $\CbL$ is an invertable matrix and $\DbL$ is a triangular
matrix. As we shall see, it turns out to be sufficient to
know $\det(\1-\zeta\DbL)$ as a function of $\zeta$ to calculate $\OL$.

Furthermore,
\beq
  \det(\1-\zeta\DbL) = \det(\1-\zeta\BbL)
\eeq
because
\beq
  (\1-\zeta\BbL)=\CbL^{-1}(\1-\zeta\DbL)\CbL~.
\eeq
Thus, we want to calculate $\det(\1-\zeta\BbL)$.

\subsection{The structure of $\BbL$}
To understand the structure of $\BbL$, we consider subspaces spanned by
$L$-cycle series indices of the type $\bigl\lbrace i, \phiC(i),
\phiI(i), \phiC\circ\phiC(i), \phiC\circ\phiI(i), \ldots\bigr\rbrace$,
containing all possible results of repeatedly applying $\phiC$ and
$\phiI$ to $i$. We call those sets {\it invariant sets of $L$-cycle
series}, which is the same as the {\it invariant sets of $L$-cycle
patterns} in \cite{Samuelsson:03}, but formulated with respect to $L$-cycle
series instead of $L$-cycle patterns. Let $\rho_L^0, \ldots,
\rho_L^{H_L-1}$ denote the invariant sets of $L$-cycle series, where
$H_L$ is the number of such sets. For convenience, let $\rho_L^0$ be
the invariant set $\lbrace 0,1\rbrace$. These definitions are
consistent with the notation in \cite{Samuelsson:03}, meaning that $H_L$ is the
same number and $\rho_0$ corresponds to the same invariant set.

The elements of $\BbL$ are given by
\beq
  B_L^{ij} = \rC\delta_{j,\phiC(i)} + \rI\delta_{j,\phiI(i)} 
                 + \delta_{i0}(c_0 - 1) + \delta_{i1}c_1~,
\eeq
which means that $\BbL$ and $\1-\zeta\BbL$ are block triangular
matrices, with blocks corresponding to the invariant sets. Consequently,
$\det(\1-\zeta\BbL)$ is the product of the determinants of all blocks
on the diagonal.

Let $r=\rC+\rI$ and $\dr=\rC-\rI$.
Then, the determinant of the block corresponding to $\rho_L^0$ is
\beq
  \det\begin{pmatrix}1-\zeta(\rC+c_0-1)&-\zeta(\rI+c_0-1)\\
                     -\zeta(\rI+c_1)&1-\zeta(\rC+c_1)\end{pmatrix} 
       = 1 - \dr\zeta~,
\eeq
because $\rC+\rI+c_0+c_1=1$.

To calculate the determinants of the blocks corresponding to the other
invariant sets, we need to explore the structure of the invariant
sets. Consider an invariant set of $L$-cycle series, $\rho_L^h$. Let
$\ell$ be the {\it length} of $\rho_L^h$, meaning that $\ell$ is the
lowest number such that, for $i \in \rho_L^h$, $(\phiC)^\ell(i)$ is
either $i$ or the index of series $i$ inverted. If $(\phiC)^\ell(i) =
i$, we say that the {\it parity} of $\rho_L^h$ is positive. Otherwise
the parity is negative. The structure of an invariant set of $L$-cycle
series is fully determined by its length and its parity. Such a set
can be enumerated on the form $\bigl\lbrace\phiC(i), \ldots,
(\phiC)^\ell(i), \phiI(i), \ldots, \phiI\circ(\phiC)^{\ell-1}(i),
\bigr\rbrace$ and for positive parity $(\phiC)^\ell(i) = i$, while
$\phiI\circ(\phiC)^{\ell-1}(i) = i$ for negative parity.

Let strings of $\T$ and $\F$ denote specific $L$-cycle series. Then
$\phiC(\F\F\F\T) = \F\F\T\F$ and $\phiI(\F\F\F\T) = \T\T\F\T$.
Examples of invariant sets of $4$-cycle series are $\lbrace\F\F\F\T$,
$\F\F\T\F$, $\F\T\F\F$, $\T\F\F\F$, $\T\T\T\F$, $\T\T\F\T$,
$\T\F\T\T$, $\F\T\T\T\rbrace$ and $\lbrace\F\T\F\T$,
$\T\F\T\F\rbrace$. The first example has length $4$ and positive
parity, while the second has length $1$ and negative parity.

Let $\Rb^+_\ell$ and $\Rb^-_\ell$ denote the blocks in
the diagonal of $\BbL$ that correspond to invariant sets of length
$\ell$ with positive and negative parity, respectively. Before we
present the form
of $\Rb^\pm_\ell$ for general $\ell$, we consider a few examples.

The invariant set $\lbrace\F\T\F\T$, $\T\F\T\F\rbrace$ corresponds
to 
\beq
  \Rb^-_1 \equiv \begin{pmatrix}\rI&\rC\\\rC&\rI\end{pmatrix}~,
\eeq
where the first row corresponds to the 4-cycle series $\F\T\F\T$
while the second row corresponds to $\T\F\T\F$. Correspondingly,
we define
\beq
  \Rb^+_1 \equiv \begin{pmatrix}\rC&\rI\\\rI&\rC\end{pmatrix}~.
\eeq
Using this definition, the other example of an invariant set,
$\lbrace\F\F\F\T$, $\F\F\T\F$, $\F\T\F\F$, $\T\F\F\F$, $\T\T\T\F$,
$\T\T\F\T$, $\T\F\T\T$, $\F\T\T\T\rbrace$, corresponds to
\beq
  \Rb^+_4 \equiv \begin{pmatrix}
     \0&\Rb_1^+&\0&\0\\
     \0&\0&\Rb_1^+&\0\\
     \0&\0&\0&\Rb_1^+\\
     \Rb_1^+&\0&\0&\0
       \end{pmatrix}~,
\eeq
with rows and columns connected to the 4-cycle series $\F\F\F\T$,
$\T\T\T\F$, $\F\F\T\F$, $\T\T\F\T$, $\F\T\F\F$, $\T\F\T\T$, $\T\F\F\F$,
and $\F\T\T\T$, in that order.

For general $\ell$ and parity, we get
\beq
  \Rb^\pm_\ell \equiv \begin{pmatrix}
     \0&\Rb_1^+&\0&\cdots&\0\\
     \0&\0&\Rb_1^+&\cdots&\0\\
     \vdots&\vdots&\vdots&\ddots&\vdots\\
     \0&\0&\0&\cdots&\Rb_1^+\\
     \Rb_1^\pm&\0&\0&\cdots&\0
       \end{pmatrix}~,
\eeq
where order of the rows (and columns) is given by index sequences of the
type $\phiC(i), \phiI(i)$, $\ldots$, $(\phiC)^\ell(i),
\phiI\circ(\phiC)^{\ell-1}(i)$.

The unitary tranformation matrix
\beq
  \Sb_1 \equiv \frac1{\sqrt2}\begin{pmatrix}1&-1\\1&1\end{pmatrix}
\eeq
transforms $\Rb^\pm_1$ according to
\beq
  \Rbt^\pm_1 \equiv \Sb_1^\top\Rb^\pm_1\Sb_1
             = \begin{pmatrix}r&0\\0&\pm\dr\end{pmatrix}~.
\eeq
Let
\beq
  \Sb_\ell \equiv \begin{pmatrix}
     \Sb_1&\0&\cdots&\0\\
     \0&\Sb_1&\cdots&\0\\
     \vdots&\vdots&\ddots&\vdots\\
     \0&\0&\cdots&\Sb_1
   \end{pmatrix}.
\eeq
Then,
\begin{align}
  \det(&\1-\zeta\Rb^\pm_\ell)=\non\\
         &= \det[\Sb_\ell^\top(\1-\zeta\Rb^\pm_\ell)\Sb_\ell] \\
         &= \det(\1-\zeta\Sb_\ell^\top\Rb^\pm_\ell\Sb_\ell) \\
  &= \det\begin{pmatrix}
     \1&-\zeta\Rbt_1^+&\0&\cdots&\0\\
     \0&\1&-\zeta\Rbt_1^+&\cdots&\0\\
     \vdots&\vdots&\vdots&\ddots&\vdots\\
     \0&\0&\0&\cdots&-\zeta\Rbt_1^+\\
     -\zeta\Rbt_1^\pm&\0&\0&\cdots&\1
       \end{pmatrix}\\
  &= [1-(r\zeta)^\ell][1\mp(\dr\zeta)^\ell]~.
\label{eq: det(1-Rz)}
\end{align}

To calculate $\det(\1-\zeta\BbL)$, we need to find the distribution of
lengths and parities of the invariant sets of $L$-cycle series.
If an $L$-cycle series belongs to an invariant set with length $\ell$
it will be itself or itself inverted after $\ell$ timesteps, depending
on whether the invariant set is of positive or negative parity. This gives
a periodicity of $\ell$ or $2\ell$ timesteps. Hence, for
$L$-cycles, there will be invariant sets of length $\ell$ and positive
parity if $\ell|L$, whereas sets of negative parity are present if
$2\ell|L$.

If invariant sets of a specific length and parity are present,
their number is independent of $L$, because
the basic form of the series in such sets does not change with
$L$. Only the number of basic repetitions differs.  Let $J^+_\ell$ and
$J^-_\ell$ denote the numbers of invariant sets of length $\ell$ with
positive or negative parity, respectively. Then,
\begin{align}
  \det(\1-\BbL\zeta)\non
     = &\prod_{\ell|L}\,
        \bigl[1-(r\zeta)^\ell\bigr]^{J_\ell^+-\delta_{\ell1}}
        \bigl[1-(\dr\zeta)^\ell\bigr]^{J_\ell^+}
     \non\\&\times
        \prod_{2\ell|L}
        \bigl[1-(r\zeta)^\ell\bigr]^{J_\ell^-}
        \bigl[1+(\dr\zeta)^\ell\bigr]^{J_\ell^-}~,
\end{align}
where the Kronecker delta, $\delta_{\ell1}$, takes care of the special
case of the block of $\BbL$, corresponding to $\rho_L^0$.  This block
has the determinant $1-\dr\zeta$, instead of the $(1-r\zeta)(1-\dr\zeta)$
it would have if it obeyed eq.~\eqref{eq: det(1-Rz)}.

\subsection{Calculation of $\OL$ via tensor calculus}
Here, we derive how $\OL$ can be determined from
$\det(\1-\BbL\zeta)$. From eq.~\eqref{eq: OL} and eq.~\eqref{eq:
AbL}, we see that
\beq
  \OL = \!\sum_{\substack{\nb \in \Nat^m\\n=N}}\binom{N}{\nb}\prod_{i=0}^{m-1}
      [(\BbL\nb/N + \eb_0)_i]^{\n_i}~.
\label{eq: OLB}
\eeq

Eq.~\eqref{eq: OLB} can be rewritten, using $(m+m)$-dimensional tensors
such that each element is indexed by two of the vectors $\vkappa,\vlambda,
\vmu,\vnu\in\Nat^m$. Each such vector corresponds to a distribution of
$L$-cycle patterns, and $\kappa$, $\lambda$, $\mu$ and $\nu$
denote the sums of the elements in each vector. We get
\beq
  \OL = \Tr(\GbL)~,
\eeq
where
\beq
  (G_L)_\vmu^\vnu \equiv \delta_{\mu N}
    \binom{N}{\vmu}\prod_{j=0}^{m-1}[(\BbL\vnu/N + \eb_0)_j]^{\mu_j}
\eeq
and the trace operator is defined as
\beq
  \Tr(\Xb) ~\equiv \hspace{-12pt}\sum_{0\leq\mu_0,\ldots,\mu_{m-1}}
          \hspace{-12pt}X_\vmu^\vmu~,
\eeq
for an $(m+m)$-dimensional tensor $\Xb$.

For convenience, we define the operation $n^{\underline{k}}$ as
\beq
  n^\ku \equiv \!\prod_{i=0}^{k-1}(n-i)~.
\label{eq: underline pow}
\eeq
We also choose the convention to interpret $n^0$ as an empty product,
independently of the value of $n$. This means that $0^0=1$, and division
must be treated with special care. Let
\begin{align}
  M_\vkappa^\vmu 
       &\equiv \!\prod_{j=0}^{m-1}\mu_j^{\kappa_j}
\label{eq: M}
\intertext{and}
  \Mt_\vkappa^\vmu 
       &\equiv \!\prod_{j=0}^{m-1}\mu_j^{\underline{\kappa_j}}
       ~~.
\end{align}
$\Mbt$ is triangular in the sense that
$\kappa_0\le\mu_0,\ldots,\kappa_{m-1}\le\mu_{m-1}$ for all non-zero
elements $\Mt_\vkappa^\vmu$. Hence, $\Mbt$ has an inverse that obeys
$\mu_0\le\kappa_0,\ldots,\mu_{m-1}\le\kappa_{m-1}$ for all non-zero
elements $(\Mt^{-1})_\vmu^\vkappa$.

Letting $\Mbt$ act on $\GbL$ yields
\begin{align}
  \Mt_\vkappa^\vmu (G_L)_\vmu^\vnu
  =&\!
  \sum_{\mu_0=\kappa_0}^\infty\,\ldots
  \hspace{-12pt}
  \sum_{\mu_{m-1}=\kappa_{m-1}}^\infty
  \hspace{-12pt}
     \delta_{\mu N}N^{\underline{\kappa}}\binom{N-\kappa}{\vmu-\vkappa}\non\\
     &\times\prod_{j=0}^{m-1}[(\BbL\vnu/N + \eb_0)_j]^{\mu_j}\\
  =&~N^{\underline{\kappa}}\prod_{j=0}^{m-1}
         [(\BbL\vnu/N + \eb_0)_j]^{\kappa_j}\non\\
     &\times\biggl[\,\sum_{j=0}^{m-1}(\BbL\vnu/N + \eb_0)_j\biggr]^{\!N-\kappa}\\
  =&~N^{\underline{\kappa}}\prod_{j=0}^{m-1}
         [(\BbL\vnu/N + \eb_0)_j]^{\kappa_j}~,
\label{eq: MtG 0}
\end{align}
because $\sum_{j=0}^{m-1}(\BbL\x + \eb_0)_j = 1$. Eq.~\eqref{eq: MtG 0}
can be rewritten to
\begin{align}
  \Mt_\vkappa^\vmu (G_L)_\vmu^\vnu
  =&~
   \frac{N^{\underline{\kappa}}}{N^\kappa}\prod_{j=0}^{m-1}
         [(\BbL\vnu)_j+N\delta_{j0}]^{\kappa_j}\\
  =&~
   \frac{N^{\underline{\kappa}}}{N^\kappa}
         \sum_{\lambda=0}^{\kappa_0}\binom{\kappa_0}{\lambda}
         [(\BbL\vnu)_0]^{\lambda}N^{\kappa_0-\lambda}\non\\
    &\times\prod_{j=1}^{m-1}
         [(\BbL\vnu)_j]^{\kappa_j}~.
  \label{eq: MtG lin 0}
\end{align}

Let $Z_\vlambda^\vkappa(\Xb)$, for an arbitrary $m\times m$
matrix $\Xb$, denote the coefficients of the formal expansion of
$\prod_{j=0}^{m-1}[(\Xb\zb)_j]^{\lambda_j}$ in such a
way that
\beq
  \prod_{j=0}^{m-1}[(\Xb\zb)_j]^{\lambda_j}
  \,\equiv\hspace{-12pt}
  \sum_{0\leq\kappa_0,\ldots, \kappa_{m-1}}
  \hspace{-12pt}
  Z_\vlambda^\vkappa(\Xb)\prod_{j=0}^{m-1}z_j^{\kappa_j}~.
\eeq
Then, we see that the operator $\Zb$ satisfies the condition
\beq
  \Zb(\Xb\Yb)=\Zb(\Xb)\Zb(\Yb)
\eeq
for all $m\times m$ matrices $\Xb$ and $\Zb$. Furthermore, $\Zb(\Xb)$
is block diagonal for any $\Xb$, in the sense that $\kappa=\lambda$
for all nonzero elements $Z_\vlambda^\vkappa(\Xb)$.

Define $\delta_{\vnu\vmu}$ according to
\beq
  \delta_{\vnu\vmu}=\!\prod_{j=0}^{m-1}\delta_{\nu_j\vmu_j}
\eeq
for arbitrary $m$-vectors $\vnu$ and $\vmu$.
Now, eq.~\eqref{eq: MtG lin 0} can be rewritten as
\beq
  \Mbt\GbL = \Nb\Eb\Zb(\BbL)\Mb~,
\eeq
where
\beq
  N_\vlambda^\vkappa = 
            \delta_{\vlambda\vkappa}\frac{N^{\underline{\kappa}}}{N^\kappa}
\eeq
and
\beq
  E_\vkappa^\vlambda
    = 
    N^{\kappa_0-\lambda_0}\binom{\kappa_0}{\lambda_0}
    \!\prod_{j=1}^{m-1}\delta_{\kappa_j\lambda_j}~.
\eeq
The block diagonal structure of $\Zb(\Xb)$ yields
\beq
  \Nb\Zb(\Xb) = \Zb(\Xb)\Nb
\eeq
for all $\Xb$, because $\Nb$ is diagonal with diagonal elements that are
constant for constant $\kappa$. According to eq.~\eqref{eq: BbL triangle},
$\BbL$ can be written as
\beq
  \BbL=\CbL^{-1}\DbL\CbL~,
\eeq
where $\DbL$ is triangular. Thus,
\begin{align}
  \Mbt\GbL &= \Nb\Eb\Zb(\CbL^{-1}\DbL\CbL)\Mb\\
  \Zb(\CbL)\Mbt\GbL &= \Nb\EbhL\Zb(\DbL)\Zb(\CbL)\Mb\\
  \Zb(\CbL)\Mbt\GbL\Mbt^{-1}\Zb(\CbL^{-1}) &= \non\\
          =\Nb\EbhL\Zb(\DbL)&\Zb(\CbL)\Mb\Mbt^{-1}\Zb(\CbL^{-1})
  \label{eq: G-transform 0}
\end{align}
where $\EbhL = \Zb(\CbL)\Eb\Zb(\CbL^{-1})$.
Because $\Mbt^{-1}$ is triangular with the lower indicies (acting to the
left) less than or equal to the upper indicies, right multiplication
with $\Mbt^{-1}$ always yields a convergent result. Similarly,
$\Zb(\CbL)$ and $\Zb(\CbL^{-1})$ are well-behaved with respect to both
left and right multiplication, as a consequence of their block
diagonal structure. Thus, both sides of the equality in eq.~\ref{eq:
G-transform 0} are well-defined.

Let $\TbL=\Zb(\CbL)\Mb\Mbt^{-1}\Zb(\CbL^{-1})$, which yields 
\beq
   \Zb(\CbL)\Mbt\GbL\Mbt^{-1}\Zb(\CbL^{-1}) = \Nb\EbhL\Zb(\DbL)\TbL~.
\eeq
The tensor $\Mb\Mbt^{-1}$ tells how to express moments in terms of
combinatorial moments. Each moment can be expressed as a sum of the
combinatorial moment of the same order and a linear combination of
combinatorial moments of lower order. Hence,
\beq
  M_\vkappa^\vmu
  (\Mt^{-1})_\vmu^\vlambda
  = \delta_{\vkappa\vlambda}
   \trm{ ~~~~~for $\lambda\geq\kappa$ .} 
\eeq

This property is conserved as $\Mb\Mbt^{-1}$ is transformed by $\Zb(\CbL)$,
because
\beq
  Z_\vnu^\vkappa(\CbL) = 0
   \trm{ ~~~~~for $\kappa\ne\nu$}~,
\eeq
meaning that
\beq
  T_\vnu^\vmu
  = \delta_{\vnu\vmu}
   \trm{ ~~~~~for $\mu\geq\nu$ .} 
  \label{eq: T triangular}
\eeq
This is also true for $\EbhL$, i.e.,
\beq
  \Eh_\vkappa^\vlambda = E_\vkappa^\vlambda
  = \delta_{\vkappa\vlambda}
   \trm{ ~~~~~for $\lambda\geq\kappa$ .} 
  \label{eq: E triangular}
\eeq

Because $\TbL$ and $\EbhL$ are triangular (with the same orientation) with
unitary block diagonal, and $\Nb$ and $\Zb(\DbL)$ are block diagonal, we get
\beq
  \OL = \Tr(\GbL) = \Tr[\Nb\EbhL\Zb(\DbL)\TbL] = \Tr[\Nb\Zb(\DbL)]~.
\eeq

The triangular structure of $\DbL$ yields that the diagonal elements
of $\Zb(\DbL)$ are given by
\beq
  Z_\vkappa^\vkappa(\DbL)=\!\prod_{j=0}^{m-1}(D_L^{jj})^{\kappa_j}~.
\eeq
Thus, 
\beq
  \Tr[\Nb\Zb(\DbL)] ~= \hspace{-12pt}
      \sum_{0\leq\kappa_0,\ldots,\kappa_{m-1}}
        \hspace{-12pt}
         \frac{N^{\underline{\kappa}}}{N^\kappa}
         \prod_{j=0}^{m-1}D_{jj}^{\kappa_j}
\eeq
and
\begin{align}
  \OL &= \sum_{\nu=0}^N\frac{N^{\underline{\nu}}}{N^\nu}
        \hspace{-4pt}
         \sum_{0\leq\kappa_0,\ldots,\kappa_{m-1}}
        \hspace{-10pt}
         \delta_{\kappa\nu}
         \prod_{j=0}^{m-1}(D_L^{jj})^{\kappa_j}\\
      &= \sum_{\nu=0}^N\frac{N^{\underline{\nu}}}{N^\nu}
      	  \frac1{\nu!}\frac{d^\nu}{d\zeta^\nu}
            \vphantom{\prod_{j=0}^{m-1}}
         \bigg|_{\zeta=0}
         ~\sum_{\substack{0\leq\kappa_0,\\\ldots,\\\kappa_{m-1}}}~
         \prod_{j=0}^{m-1}(D_L^{jj}\zeta)^{\kappa_j}\\
      &= \sum_{\nu=0}^N\frac1{N^\nu}\binom N\nu
      	  \frac{d^\nu}{d\zeta^\nu}
            \vphantom{\prod_{j=0}^{m-1}}
         \bigg|_{\zeta=0}
      	  \prod_{j=0}^{m-1}\frac1{1-D_L^{jj}\zeta}\\
      &= \biggl(1+\frac1N\frac d{d\zeta}\biggr)^{\!\!N}\bigg|_{\zeta=0}
          \prod_{j=0}^{m-1}\frac1{1-D_L^{jj}\zeta}~.
\end{align}
Because
\begin{align}
  \prod_{j=0}^{m-1}(1-D_L^{jj}\zeta) 
     &= \det(\1-\DbL\zeta)\\
     &= \det(\1-\BbL\zeta)~,
\end{align}
we get
\beq
  \OL = \biggl(1+\frac1N\frac d{d\zeta}\biggr)^{\!\!N}\bigg|_{\zeta=0}
        \,\frac1{\det(\1-\BbL\zeta)}~,
\eeq
where
\begin{align}
  \det(\1-\BbL\zeta)\non
     = &\,\prod_{\ell|L}
        \bigl[1-(r\zeta)^\ell\bigr]^{J_\ell^+-\delta_{\ell1}}
        \bigl[1-(\dr\zeta)^\ell\bigr]^{J_\ell^+}
     \non\\&\times
        \prod_{2\ell|L}
        \bigl[1-(r\zeta)^\ell\bigr]^{J_\ell^-}
        \bigl[1+(\dr\zeta)^\ell\bigr]^{J_\ell^-}~.
\label{eq: det(1-Bz) 0}
\end{align}

\subsection{Calculation of $J_\ell^\pm$}
Let $\psi_\ell^+$ and
$\psi_\ell^-$, respectively, denote the sets of (infinite) time series of
$\TRUE$ and $\FALSE$, such that each series is identical to, or the
inverse of, itself after $\ell$ time steps. Then, the set of time
series that are part of an invariant set of $L$-cycles with length
$\ell$ and negative parity, ${\cal J}^-_\ell$, is given by
\beq
  {\cal J}^-_\ell = \,\psi^-_\ell \,\big\backslash
       \hspace{-10pt}
       \bigcup_{\substack{d\textrm{ odd prime}\\d|\ell}}
       \hspace{-10pt}
       \psi^-_{\ell/d}~.
\eeq
For positive parity, we get
\beq
  {\cal J}^+_\ell = \,\psi^+_\ell \bigg\backslash
       \Biggl(
         \bigcup_{\substack{d\textrm{ prime}\\d|\ell}}\! \psi^+_{\ell/d}
       \Biggr)
       \bigg\backslash {\cal J}^-_{\ell/2}~,
\eeq
where ${\cal J}^-_{\ell/2}$ is the empty set if $\ell/2$ is not an integer.

The numbers of elements in ${\cal J}^\pm_\ell$ are given by $2\ell J^\pm_\ell$,
where $J^\pm_\ell$ are the numbers of invariant sets with length $\ell$.
Then, the inclusion--exclusion principle yields
\beq
  J^-_\ell = \,\frac1{2\ell}\hspace{-6pt}
     \sum_{~\mathbf{s}\in\lbrace0,1\rbrace^{\tilde{\eta}_\ell}}
           \!\!
         (-1)^s2^{\ell/\tilde{d}_\ell(\mathbf{s})}~,
  \label{eq: Jl-}
\eeq
where $s = \sum_{i=1}^{\tilde{\eta}_\ell} s_i$, $\tilde{d}(\mathbf{s}) =
\prod_{i=1}^{\tilde{\eta}_\ell} (\tilde{d}_{\ell}^i)^{s_i}$ and
$\tilde{d}_\ell^1,\ldots,\tilde{d}_\ell^{\eta_{\ell}}$ 
are the odd prime divisors to $\ell$. Similarly
\beq
  J^+_\ell = \,\tfrac1{2\ell}\hspace{-8pt}
         \sum_{\substack{s_0\in\lbrace0,1\rbrace\\2^{s_0}|\ell}}\!\!
         \sum_{~\mathbf{s}\in\lbrace0,1\rbrace^{\tilde{\eta}_\ell}}
         \hspace{-8pt}
         (-1)^{s_0+s}2^{\ell/d_\ell(s_0,\mathbf{s})}-\tfrac12J^-_{\ell/2}~,
  \label{eq: Jl+0}
\eeq
where $d(s_0,\mathbf{s})=2^{s_0}\tilde{d}(\mathbf{s})$ and $J^-_{\ell/2}=0$
if $\ell/2$ is not an integer. Insertion of Eq.~\ref{eq: Jl-} into
Eq.~\ref{eq: Jl+0} yields
\beq
  J^+_\ell = \,\tfrac1{2\ell}\hspace{-8pt}
         \sum_{\substack{s_0\in\lbrace0,1\rbrace\\2^{s_0}|\ell}}\!\!
         \sum_{~\mathbf{s}\in\lbrace0,1\rbrace^{\tilde{\eta}_\ell}}
         \hspace{-8pt}
         (1+s_0)(-1)^{s_0+s}2^{\ell/d_\ell(s_0,\mathbf{s})}~,
  \label{eq: Jl+}
\eeq
which also can be written as
\beq
  J^+_\ell = J^-_\ell - J^-_{\ell/2}~.
\label{eq: J+ from J-}
\eeq

\subsection{Calculation of $\OL$ for small and large $N$}

Iterative application of eq.~\eqref{eq: J+ from J-} and the identity
\beq
  [1-(\dr\zeta)^\ell][1+(\dr\zeta)^\ell] = 1-(\dr\zeta)^{2\ell}
\eeq
to eq.~\eqref{eq: det(1-Bz) 0} yields
\begin{align}
  \det(&\1-\BbL\zeta)=\non\\
     =& \hspace{-4pt}
        \prod_{\substack{\ell|L\\L/\ell\textrm{ odd}}}
        \hspace{-4pt}
        \bigl[1-(r\zeta)^\ell\bigr]^{J_\ell^--J_{\ell/2}^--\delta_{\ell1}}
        \bigl[1-(\dr\zeta)^\ell\bigr]^{J_\ell^-}
     \non\\&\times
        \hspace{-6pt}
        \prod_{\substack{\ell|L\\L/\ell\textrm{ even}}}
        \hspace{-4pt}
        \bigl[1-(r\zeta)^\ell\bigr]^{2J_\ell^--J_{\ell/2}^--\delta_{\ell1}}
\label{eq: det(1-Bz) 1}~,
\end{align}
where $J^-_\ell$ is given by eq.~\eqref{eq: Jl-}.

The simplest way to calculate $\OL$ is to work with power series
expansions. If
\beq
  \sum_{k=0}^\infty c_k\zeta^k=\frac1{\det(\1-\BbL\zeta)}~,
\label{eq: z powers}
\eeq
$\OL$ is given by
\beq
  \OL = \sum_{k=0}^N\frac{N^{\underline{k}}}{N^k}c_k~,
\label{eq: OL of z powers}
\eeq
where the operation $N^{\underline{k}}$ is defined in
eq.~\eqref{eq: underline pow}.

For $r<1$, $\zeta=1$ is within the convergence radius of the power
series in eq.~\eqref{eq: z powers}. Hence, we get
\beq
  \lim_{N\rightarrow\infty}
    \biggl(1+\frac1N\frac d{d\zeta}\biggr)^{\!\!N}\bigg|_{\zeta=0}
          \,\frac1{\det(\1-\BbL\zeta)}
  = \frac1{\det(\1-\BbL)}~,
\eeq
which means that the large $N$ limit of $\OL$ is given by
\beq
  \OLinf = \frac1{\det(\1-\BbL)}~.
\label{eq: OLinf}
\eeq
This result is consistent with the calculations in \cite{Kauffman:04},
which also are valid for subcritical networks with many inputs per node.

For $r=1$, the dominant contribution to $\OL$ comes from the pole at
$\zeta=1$. For $|\dr| < 1$ and large $N$, we get
\beq
  \OL \approx \biggl(1+\frac1N\frac d{d\zeta}\biggr)^{\!\!N}\bigg|_{\zeta=0}
      \,\frac{\gamma_L}{(1-\zeta)^{H_L-1}}
\label{eq: OL crit high N}
\eeq
where $H_L$ is the total number of invariant $L$-cycle sets and $\gamma_L$
is a constant. For large $N$, a power series expansion of
$1/[(1-\zeta)^{H_L-1}]$ yields
\begin{align}
  \OL &\approx \gamma_L
            \sum_{k=0}^N\frac{N^{\underline{k}}}{N^k}\binom{H_L-2+k}{k}
\label{eq: OL crit expansion}\\
      &\approx \tilde{\gamma}_L\int_0^\infty\!\!dk\,k^{H_L-2}e^{-k^2/N}~,
\label{eq: OL crit integral}
\end{align}
where $\tilde{\gamma}_L$ is a constant. The dominant terms in
eq.~\eqref{eq: OL crit expansion} satisfy $H_L\ll k\ll N$ in the
large $N$ limit, and Stirling's formula has been applied to those.
From eq.~\eqref{eq: OL crit integral}, we see that the asymptotic
behavior of $\OL$ is the scaling
\beq
  \OL \propto N^{(H_L-1)/2}~,
\label{eq: OL power 0}
\eeq
for critical $r=1$ networks with $|\dr| < 1$. This also means that
\beq
  \CL \propto N^{(H_L-1)/2}
\label{eq: CL power 0}
\eeq
for large $N$.

If $|\dr|=1$, the network only consists of copy operators (if
$\dr=1$) or inverters (if $\dr=-1$). Then, $H_L$ should be replaced
by the number of sets of $L$-cycle series that are invariant under
the present choice of operators. This number will be as least as large as
$H_L$, because the invariant set can split, but not merge, when one
operator is removed.

One can also use complex analysis to retrieve the operator identity
\beq
  \biggl(1+\frac1N\frac d{d\zeta}\biggr)^{\!\!N}\bigg|_{\zeta=0}
      \hspace{-10pt}[\cdots]
    = \frac{N!}{2\pi iN^N}\!\oint_{|\zeta|=\epsilon}
      \hspace{-8pt}d\zeta
        \frac{e^{N\zeta}}{\zeta^{N+1}}[\cdots]~,
\eeq
where $\epsilon$ is a positive constant small enough to keep
all poles of the given function outside $|\zeta|\leq\epsilon$.  This
is especially useful to calculate the total number of states in
attractors; $\OL[0]$.

Eq.~\eqref{eq: det(1-Bz) 1} can be rewritten as the exponential of a
power series expansion. The expansion
\beq
  -\ln(1 - z) = \sum_{j=1}^\infty\frac{x^j}j
\eeq
yields
\begin{align}
  -\ln\det(&\1-\BbL\zeta)=\non\\
     =& \hspace{-4pt}
        \sum_{\substack{\ell|L\\L/\ell\textrm{ odd}}}
        \hspace{-4pt}
        \biggl[
          J^{\trm{odd}}_\ell\ell
              \sum_{j=1}^\infty\frac{(r\zeta)^{j\ell}}{j\ell}
         + J_\ell^-\ell
              \sum_{j=1}^\infty\frac{(\dr\zeta)^{j\ell}}{j\ell}
        \biggr]
     \non\\&+
        \hspace{-6pt}
        \sum_{\substack{\ell|L\\L/\ell\textrm{ even}}}
        \hspace{-4pt}
        J^{\trm{even}}_\ell\ell
              \sum_{j=1}^\infty\frac{(r\zeta)^{j\ell}}{j\ell}~,
\end{align}
where $J^{\trm{odd}}_\ell=J_\ell^--J_{\ell/2}^--\delta_{\ell1}$ and
$J^{\trm{even}}_\ell=2J_\ell^--J_{\ell/2}^--\delta_{\ell1}$.

Let $(a,b)$ denote the greatest common divisor of $a$ and $b$.
Then, the substitution $k=j\ell$ and reordered summation gives
\begin{align}
  -\ln\det(&\1-\BbL\zeta)=\non\\
     =&~\, \sum_{k=1}^\infty\frac{\zeta^k}k
         \hspace{-6pt}
        \sum_{\substack{\ell|(k,L)\\L/\ell\textrm{ odd}}}
         \hspace{-4pt}
          [r^kJ^{\trm{odd}}_\ell\ell
         + (\dr)^kJ_\ell^-\ell]
     \non\\&+\, \sum_{k=1}^\infty\frac{(r\zeta)^k}k \hspace{-6pt}
        \sum_{\substack{\ell|(k,L)\\L/\ell\textrm{ even}}}
        J^{\trm{even}}_\ell\ell\\
     =&
         \hspace{-6pt}
        \sum_{\substack{k\ge0\\L/(k,L)\textrm{ odd}}}
         \hspace{-4pt}
        \frac{\zeta^k}k
         \hspace{-4pt}
        \sum_{\substack{\ell|(k,L)\\(k,L)/\ell\textrm{ odd}}}
         \hspace{-6pt}
          [-r^k + (\dr)^k]J_\ell^-\ell
     \non\\&+\, \sum_{k=1}^\infty\frac{(r\zeta)^k}k
        \!\!\sum_{\ell|(k,L)}\!\!
        J^{\trm{even}}_\ell\ell~.
\end{align}
The inner sums can be calulated using the connection between
$J_\ell^-$ and $\psi_\ell^-$. Thus,
\begin{align}
  -\ln\det(\1-\BbL\zeta)
     =& 
         \hspace{-12pt}
     \sum_{\substack{k\ge0\\L/(k,L)\textrm{ odd}}}
         \hspace{-12pt}
       \frac{\zeta^k}k
         [-r^k + (\dr)^k]2^{(k,L)-1}
     \non\\&+ \sum_{k=1}^\infty\frac{(r\zeta)^k}k(2^{(k,L)}-1)~.
\end{align}

For $L=0$, we get
\beq
  -\ln\det(\1-\Bb_0\zeta)
     = \sum_{k=1}^\infty\biggl[\frac{(2r\zeta)^k}k
        -\frac{(r\zeta)^k}k\biggr]~,
\eeq
which means that
\beq
  \frac1{\det(\1-\Bb_0\zeta)}=\frac{1-r\zeta}{1-2r\zeta}
\label{eq: det Bb0}
\eeq
and
\beq
  \OL[0]=\frac1{2\pi iN^N}\!\oint_{|\zeta|=\epsilon}
       \hspace{-8pt}d\zeta
        \frac{e^{N\zeta}}{\zeta^{N+1}}\frac{1-r\zeta}{1-2r\zeta}~.
\label{eq: Omega oint}
\eeq

The $N$-dependent part of the integrand in eq.~\eqref{eq: Omega oint}
has a saddle point at $\zeta = 1$. If $r<1/2$, the path of steepest
descent integration does not include any other poles than the desired
one at $\zeta=0$. On the other hand, if $r>1/2$ the pole at
$\zeta=1/(2r)$ gives another contribution to $\OL[0]$. The borderline
case, $r=1/2$, is similar to the case $r=1$ for $\OL$ with $L\ne0$.

Considering the properties mentioned above, we find that
the asymptotic behavior of $\OL[0]$ is given by
\beq
  \OL[0] \approx \left\{\begin{array}{ll}
             \vphantom{\bigg|}\dst
                  \frac{1-r}{1-2r}&\trm{for }r<1/2\\
             \vphantom{\bigg|}
                 \frac12\sqrt{\frac\pi2N}\dst&\trm{for }r=1/2\\
             \vphantom{\bigg|}
                 \sqrt{\frac\pi2N}e^{(\ln2r-1+1/2r)N}\dst&\trm{for }r>1/2
             \end{array}\right.~.
\label{eq: OL0 high N}
\eeq

\section{Results}

\begin{figure}[tbf]
\begin{center}
\epsfig{figure=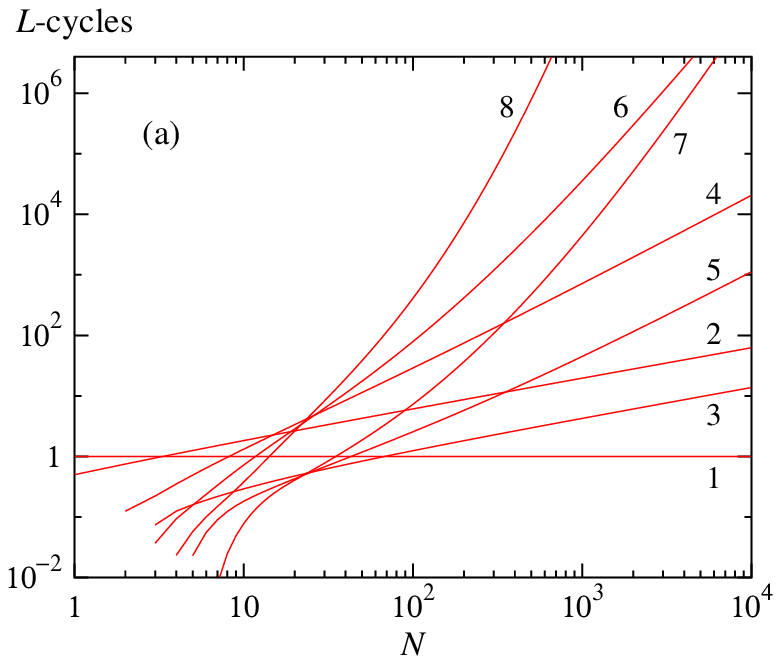}
\epsfig{figure=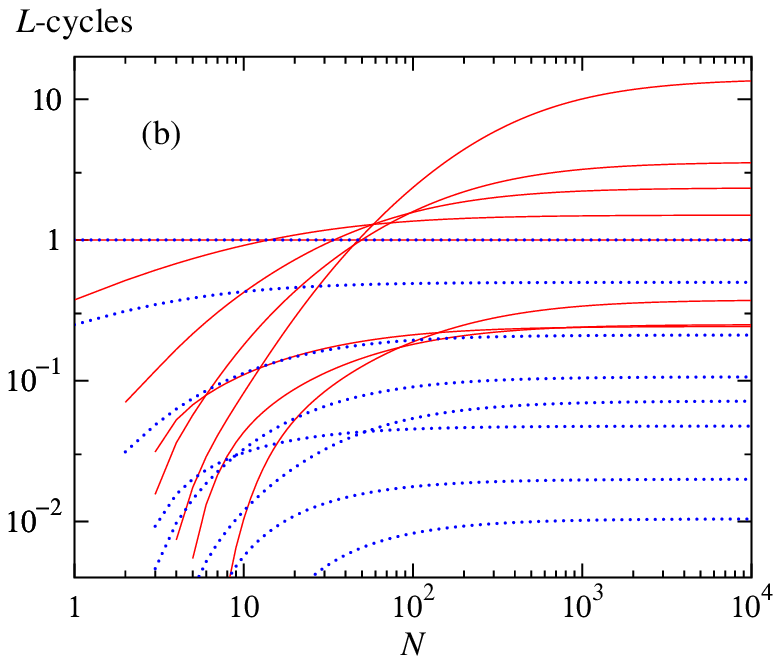}
\end{center}
\caption{The average number of proper $L$-cycles as a function of $N$
for different $L$, for networks with single-input nodes. $r=1$ in (a),
and $r=3/4$ (solid lines) and $r=1/2$ (dotted lines) in (b). In (a),
$L$ is indicated in the plot. In (b), $L$ is 3, 5, 7, 1, 2, 4, 6, 8
for $r=3/4$ and 7, 5, 3, 8, 6, 4, 2, 1 for $r=1/2$, in that order, from
bottom to top along the right boundary of the plot area. In (b), the
curves for $L=3$ and $L=5$ for $r=3/4$
essentially coincide at the right side
of the plot, whereas they split up to the left, with $L=3$ as the upper
curve there.}
\label{fig: L cyc sym}
\end{figure}

\begin{figure}[tbf]
\begin{center}
\epsfig{figure=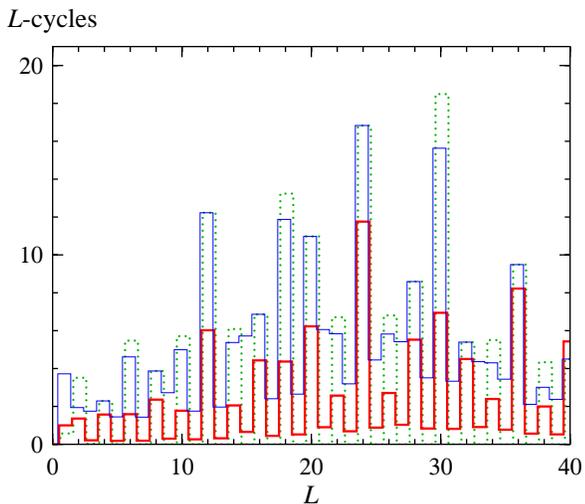}
\end{center}
\caption{The average number of proper $L$-cycles for networks with $N=100$ and
$r=3/4$, as function of $L$. $\dr=3/4$ (thin solid line), $\dr=0$
(thick solid line) and
$\dr=-3/4$ (dotted line). Note the importance of what numbers divide
$L$.}
\label{fig: L cyc asym}
\end{figure}

Our main results from the analytical calculations are the expressions
that yield $\OL$ and $\CL$ for finite $N$, and their asympotic growth.
See eqs.~\eqref{eq: CL of OL},
(\ref{eq: det(1-Bz) 1}--\ref{eq: OL of z powers}) and
\eqref{eq: det Bb0} on expressions for general $N$, and
eqs.\ \eqref{eq: OLinf}, \eqref{eq: OL power 0}, \eqref{eq: CL power 0}
and \eqref{eq: OL0 high N} on expressions valid for the high $N$ limit.
Fig.~\ref{fig: L cyc sym} shows the numbers of attractors with short
length as a function of system size, for different $r$, but with $\dr=0$.
For critical networks, with $r=1$, the asymptotic growth of $\CL$ is a
power law, while $\CL$ approaches a
constant for subcritical networks as $N$ goes to infinity.

For networks with $\dr\ne0$, the prevalences of copy operators and
inverters are not the same. Cycles of even length are in general more
prevalent then cycles of odd length. An increased number of inverters
strengthens this difference, while a low fraction of inverters makes
the difference less pronounced. See fig.~\ref{fig: L cyc asym}, which
shows the symmetric case $\dr=0$ and the extreme cases $\dr=\pm r$.

\begin{figure}[tbf]
\begin{center}
\epsfig{figure=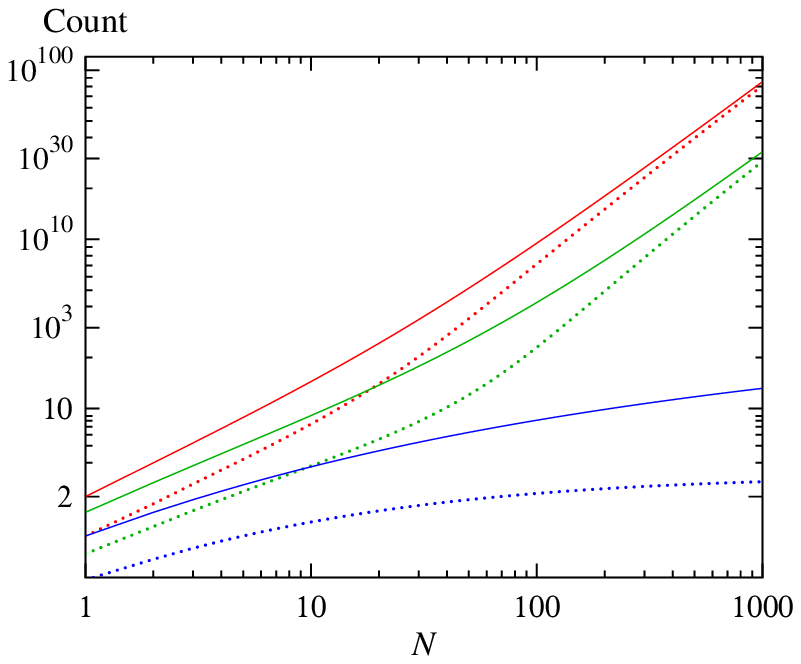}
\end{center}
\caption{$\OL[0]$ (solid lines) and $\CL[]$ (dotted lines) for
$r=1/2$, $r=3/4$ and $r=1$, in that order, from the bottom to the top
of the plot. $\dr=0$ in all cases. Note that $\OL[0]$ is independent
of $\dr$. Due to the double logarithmic, $\CL$ appears to approach
$\OL$ for $r=3/4$ and $r=1$. This is not the case --- have in mind
that a constant translation in a double logarithmic scale corresponds
to a rise to a constant power.}
\label{fig: Omega0 and C}
\end{figure}

\begin{figure}[tbf]
\begin{center}
\epsfig{figure=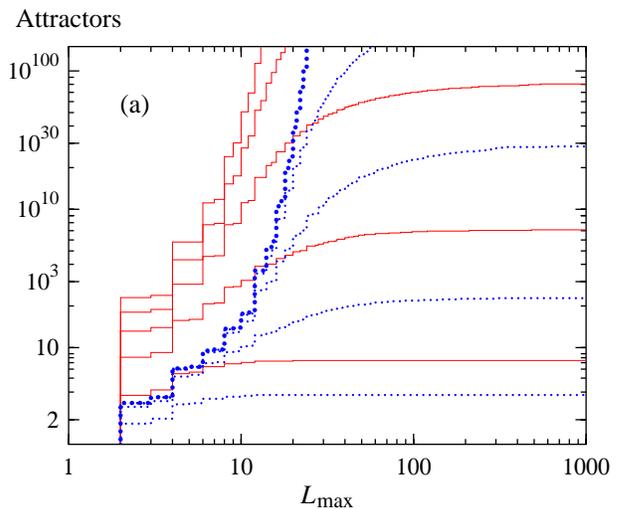}
\epsfig{figure=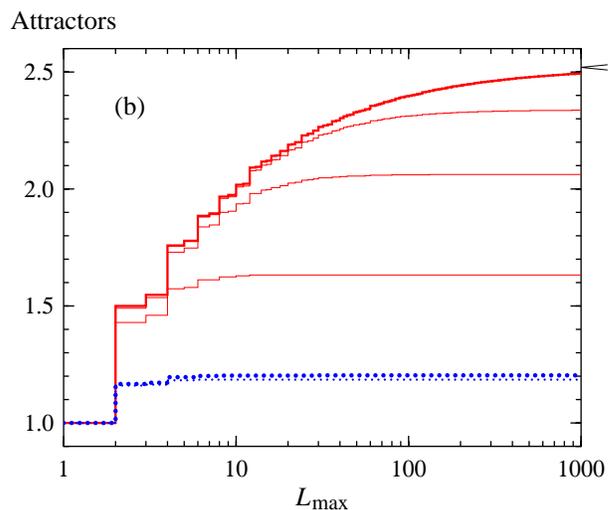}
\end{center}
\caption{The number of attractors with lengths $L\le L_{\trm{max}}$ in
networks with $N$ single-input nodes, for different values of $N$.
In (a) $N=10,10^2,\ldots, 10^5$
for $r=1$ (thin solid lines) and $N=10,\ldots,10^4$ for $r=3/4$ (thin
dotted lines). In (b) $N=10,10^2,10^3$ (thin solid lines) for $r=1/2$
and $N=10$ for $r=1/4$ (thin dotted line). For all cases,
$\dr=0$. The thick lines in (a) and (b) 
show the limiting number of attractors when
$N\rightarrow\infty$. The arrowhead in (b) marks this limit for
$L_{\trm{max}}=10^7$ for $r=1/2$. The small increase in the number of
attractors when $L_{\trm{max}}$ is changed from $10^3$ to $10^7$
indicates that $\CL[]$ converges when $N\rightarrow\infty$. Note the
drastic change in the $y$-scale between the case $r>1/2$ and
$r\le1/2$.}
\label{fig: cum cyc}
\end{figure}

The total number of attractors, $\CL[]$, and the total number of
states in attractors, $\OL[0]$, can diverge for large $N$, even though the
number of attractors of any fixed length converges. This is true for
subcritical networks with $r > 1/2$.
See figs.~\ref{fig: Omega0 and C} and \ref{fig: cum cyc}a.
The growth of $\OL[0]$ is
exponential if $r > 1/2$. Interestingly, there is no qualitative
difference in the growth of $\OL[0]$ when comparing the critical case,
$r = 1$, to the subcritical ones with $1 > r > 1/2$.

For $r < 1/2$, both $\CL[]$ and $\OL[0]$ converge to constants for
large $N$. In the border line case $r=1/2$, $\OL[0]$ diverges like a
square root of $N$, but $\CL[]$ seems to approach a constant. See
fig.~\ref{fig: cum cyc}b.

The quantity $\OL[0]$ has an interesting graph theoretical
interpretation. Let $N_{\trm{active}}$ be the fraction of
nodes that are part of a loop of non-constant nodes. Then,
\beq
  \OL[0]=\langle\exp(N_{\trm{active}}\ln 2)\rangle
\eeq

which means that $\ln\langle\exp(N_{\trm{active}})\rangle$
grows linearly with $N$ if $1/2<r\leq1$. This stands in sharp contrast
to $\langle N_{\trm{active}}\rangle$, which grows like
$\sqrt{N}$ for $r=1$ and approaches a constant for $r<1$ as
$N\rightarrow\infty$, see \cite{Flyvbjerg:88}. This means that the
distibution of $N_{\trm{active}}$ has a broad tail if
$r>1/2$.

Like $\OL[0]$ is connected to the the number of nodes in active loops,
$\CL[1]$ for $\dr=r$ is connected to the number of active loops. In
this case, however, there is no striking difference between the
average number of fixed points and the same quantity for a typical
network.

\begin{figure}[tbf]
\begin{center}
\epsfig{figure=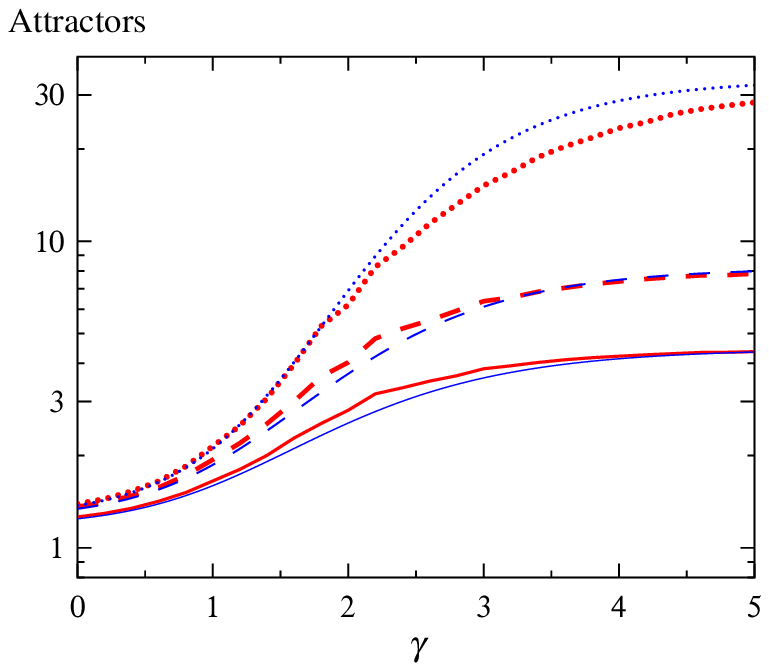}
\end{center}
\caption{Comparison between simulations for power law in-degree
networks of size $N=20$ (bold lines) and the corresponding networks
with single-input nodes (thin lines). The fitted networks have
identical values for $r$, $\dr$, and $N$. The solid lines show the
number of fixed points, whereas the dashed and dotted lines show the
numbers of 2-cycles plus fixed points and the total numbers of
attractors, respectively. The probability distribution of in-degrees
satisfies $p_K\propto K^{-\gamma}$, where $K$ is the number of inputs.
The power law networks use the nested
canalyzing rule distribution presented in \cite{Kauffman:04}.}
\label{fig: power vs 1inp}
\end{figure}

All the properties above are derived and calculated for networks with
one input per node, but they seem to a large extent to be valid for
networks with multi-input nodes. For such networks, we define $r$ and
$\dr$ as measures of disturbance propagation according to the
following procedure:\\
Find the mean field equilibrium fraction of nodes that have the value
$\TRUE$. Pick a random state from this equilibrium as an initial
configuration. Let the system evolve one time step, with and
without first toggling the value of a randomly selected node.
The average fraction of nodes that in both cases copy or invert the
state of the selected node are $\rC$ and $\rI$, respectively,
Finally, let $r=\rC+\rI$ and $\dr=\rC-\rI$. For a more detailed
description, see \cite{Kauffman:04}.

From \cite{Kauffman:04}, we know that for subcritical networks the limit of
$\CL$ as $N\rightarrow\infty$ is only dependent on $r$ and $\dr$.
Hence, we can expect that $\CL$ for a subcritical network with
multi-input nodes can be approximated with $\CL'$, calculated for a network
with single-input nodes, but with the same $r$ and $\dr$. For the
networks in \cite{Kauffman:04}, with a power law in-degree distribution, this
approximation fits surprisingly well. See fig.~\ref{fig: power vs 1inp}.

For the critical Kauffman model with in-degree 2, one can do a similar
comparison. The number of nodes that are non-constant grows
like $N^{2/3}$ for large $N$. See \cite{Socolar:03, Samuelsson:03}.
Furthermore, the effective connectivity between the non-constant
nodes approaches 1 for large $N$ \cite{Bastolla:98}.
Hence, one can expect that this type of $N$-node Kauffman networks can be
mimicked by networks with $N'=N^{2/3}$ 1-input nodes. For those
networks, we choose $r=1$ and $\dr=0$, which are the same values as for
the Kauffman networks.

\begin{figure}[tbf]
\begin{center}
\epsfig{figure=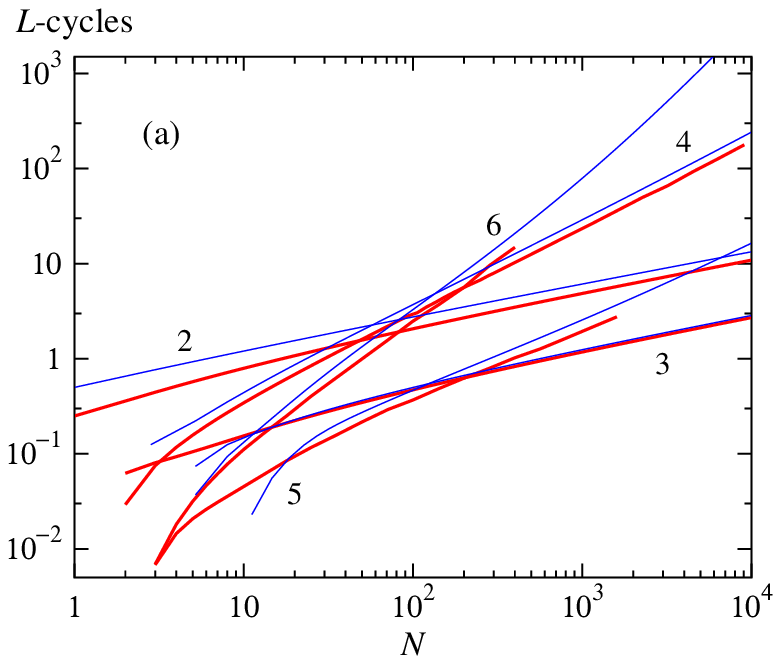}
\epsfig{figure=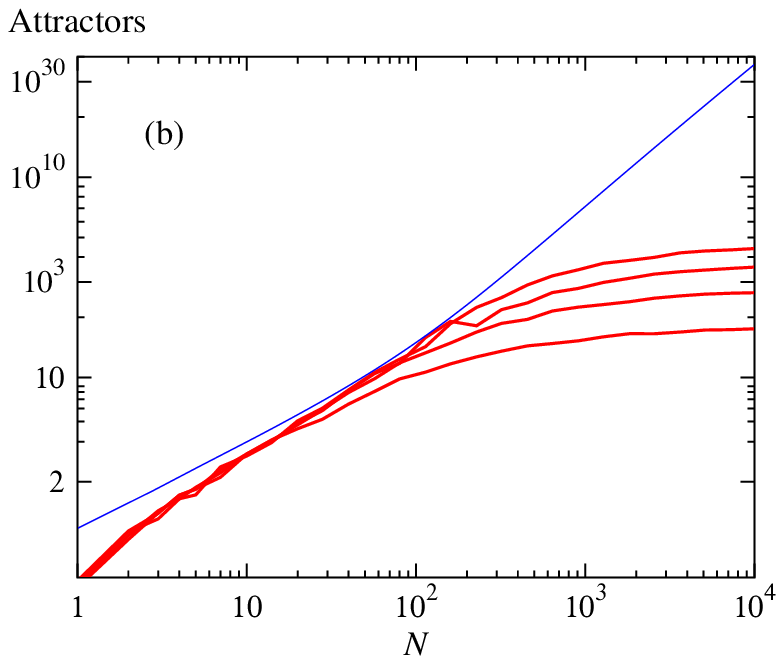}
\end{center}
\caption{Comparison between critical $K=2$ Kauffman networks (thick
lines) and the corresponding networks of single-input nodes (thin
lines). The size of the single-input networks is set to
$N'=N^{2/3}$. $r=1$ and $\dr=0$, consistent with the Kauffman
model. (a) The number of proper $L$-cycles for the $L$ indicated in the plot.
For the Kauffman networks, the
numbers have been calculated from Monte Carlo summation
for those network sizes where could could not be calculated exactly. See
\cite{Samuelsson:03}. The number of fixed points is $1$, independently
of $N$, for both network types. (b) Total
number of attractors. This quantity has been estimated by simulations
for the Kauffman networks, using $10^2$, $10^3$, $10^4$ and $10^5$
random starting configurations per network.}
\label{fig: K2 vs 1inp}
\end{figure}

For large $N$, $\CL$ in the Kauffman networks grows like
$N^{(H_L-1)/3}$, where $H_L$ is the number of invariant $L$-cycle
sets. For the selected networks with 1-input nodes, we have $\CL'
\propto N'^{(H_L-1)/2} \propto N^{(H_L-1)/3}$ for large $N'$, see
eq.~\ref{eq: CL power 0}. This confirms that the choice $N'=N^{2/3}$
is reasonable, but it does not indicate whether the proportionality
factor in $N'\propto N^{2/3}$ is close to 1. This factor could also be
dependent on $L$, as can be seen from the calculations in
\cite{Samuelsson:03}. However, this initial guess turns out to
be surprisingly good, see fig.~\ref{fig: K2 vs 1inp}a.

From the good agreement for short cycles, one can expect a similar
agreement in the total number of attractors. This is investigated in
fig.~\ref{fig: K2 vs 1inp}.
For networks with up to about 100 nodes, the agreement is
good, and the extremely fast growth of $\CL'$ for larger $N$ is
consistent with the slow convergence in the simulations.

\section{Summary and Discussion}

Using analytical tools, we have investigated random Boolean networks
with single-input nodes. We extract the exact distributions of cycles
with lengths up to 1000 in networks with up to $10^5$ nodes.
Furthermore, we find some interesting scaling properties that hold
for large $N$.

As has been pointed out in earlier work \cite{Bastolla:97},
we see that a small
fraction of the networks have many more cycles than a typical
network. This property becomes more pronounced as the system size
grows, and changes the scaling of the average number of states belonging to
cycles drastically. For networks that have the stability parameter
$r>1/2$, the average number of states in attractors grows
exponentially with the system size $N$, whereas this number grows
exponentially with $\sqrt N$ for a typical network if $r=1$, and
approaches a constant if $r<1$.

The dynamics in random Boolean networks with multi-input nodes can to
a large extent be understood in terms of the simpler single-input case.
In a direct comparison to critical Kauffman networks of connectivity
two and to subcritical networks with power law in-degree, the agreement
is surprisingly good.

Our results highlight some previously observed artefacts in random
Boolean networks. The synchronous updates lead to dynamics that
largely is governed by integer divisibility effects. Furthermore,
when counting attractors in large networks, most of them are found in
highly atypical networks and have attractor basins that are extremely
small compared to the full state space.

In \cite{Klemm:04}, a new concept of stability in attractors of Boolean
networks is presented. To only consider that type of stable attractors
is one way to make more relevant comparisons to real systems. Another
way is to focus on fixed points and stability properties as in
\cite{Kauffman:03, Kauffman:04}.
Furthermore, the limit of large systems may not always make
sense in comparison with real systems. Small Boolean networks may tell
more about real systems than large networks would.

Although there are problems in making direct comparisons between random
Boolean networks and real systems, we think that insight in the
dynamics of Boolean networks will improve the general understanding of
complex systems.  For example, can real systems have lots of
attractors that are never found due to small attractor basins, and
what implications would such attractors have on the system?

Another interesting viewpoint is to compare with combinatorial
optimization problems, e.g.\ scheduling and digital circuit design.
The question of finding an
attractor in a random Boolean network is similar to such problems.
For those problems, an attractor with a small attractor basin
corresponds to a solution that is hard to find. Carrying the analogy
further, we should not be surprised if a combinatorial problem that
seem to be impossible to solve has plenty of solutions. Random Boolean
networks could provide a playground for algorithms that handle such
problems.

\end{document}